\documentclass{article}
\usepackage[usenames,dvipsnames,svgnames]{xcolor}
\usepackage{hyperref}
\hypersetup{
    colorlinks=true,
    citecolor=MidnightBlue,
    urlcolor=Bittersweet,
}

\usepackage[margin=1in]{geometry}
\usepackage{algorithm,algpseudocode} % algorithmicx package

\usepackage{float}
\usepackage{bm}
\usepackage{extarrows}
\usepackage{comment}
\usepackage{amsthm,amsmath,amsfonts}
\usepackage{booktabs}
\usepackage{cleveref}

\usepackage{graphicx}
\usepackage{bbm}
\newcommand{\R}{\mathbb{R}}
\newcommand{\C}{\mathbb{C}}

\newcommand{\x}{\mathbf{x}}
\newcommand{\y}{\mathbf{y}}
\newcommand{\z}{\mathbf{z}}

\newcommand{\A}{\mathbf{A}}

\usepackage{tikz}

\usepackage[htt]{hyphenat}
\usepackage{xspace}
\usepackage{mathtools}
\usepackage{subcaption}
\usepackage[normalem]{ulem}

\usepackage{array}
\usepackage{textcomp}
\usepackage{stfloats}
\usepackage{url}
\usepackage{verbatim}
\usepackage{cite}
\usepackage{breakcites}

\newcommand{\M}{\mathbf{M}}
\newcommand{\fft}{\mathcal{F}}

\newcommand{\mnet}{\texttt{MNet}}
\newcommand{\mb}{\texttt{mask-backward}}

\newcommand{\del}[1]{\textcolor{blue}{\sout{#1}}}
\newcommand{\add}[1]{{\textcolor{red}{#1}}}

\newenvironment{edits}{}{}

\excludecomment{edits}

\begin{document}
\title{Single-pass Object-adaptive Data Undersampling and Reconstruction for MRI\footnote{© 2022 IEEE. Personal use of this material is permitted. Permission from IEEE must be obtained for all other uses, in any current or future media, including reprinting/republishing this material for advertising or promotional purposes, creating new collective works, for resale or redistribution to servers or lists, or reuse of any copyrighted component of this work in other works. This article is accepted for publication by IEEE with DOI: \href{https://ieeexplore.ieee.org/document/9757874}{10.1109/TCI.2022.3167454}}}
%An end-to-end object-adaptive deep network for undersampled data acquisition in MRI
%A deep learning approach for adaptive sampling with application to MRI}
\author{Zhishen Huang\thanks{Z. Huang is with the Department of Computational Mathematics, Science and Engineering, Michigan State University, East Lansing, MI, 48824 (huangz78@msu.edu).} , % <-this % stops a space
Saiprasad Ravishankar\thanks{S. Ravishankar is with the Department of Computational Mathematics, Science and Engineering and the Department of Biomedical Engineering, Michigan State University, East Lansing, MI, 48824 (ravisha3@msu.edu).
This work was supported in part by a research gift from the Advanced Radiology Services (ARS) Foundation.
}}

\maketitle

\begin{abstract}
%Accelerating the in acquisition process in MRI is realized by adopting a particular sampling scheme to undersample information in the frequency domain. 
There is recent interest in techniques to accelerate the data acquisition process in MRI by acquiring limited measurements. 
Sophisticated reconstruction algorithms are often deployed to maintain high image quality in such settings.
In this work, we propose a data-driven sampler using a convolutional neural network, \mnet, to provide object-specific sampling patterns adaptive to each scanned object. The network observes limited low-frequency $k$-space data for each object and predicts the desired undersampling pattern in one go that achieves high image reconstruction quality.
%outputs the corresponding undersampling pattern for high-frequency domain. 
We propose an accompanying alternating-type training framework 
\begin{edits}
\del{with a \mb{} procedure}
\end{edits}
that efficiently generates training labels for the sampler network and jointly trains an image reconstruction network. Experimental results on the fastMRI knee dataset demonstrate the capability of the proposed learned undersampling network to generate object-specific masks at fourfold and eightfold acceleration that achieve superior image reconstruction performance than several existing schemes.
%and the performance of corresponding image reconstruction based on information acquired according to the adaptive sampling scheme. 
The source code for the proposed joint sampling and reconstruction learning framework is available at \url{https://github.com/zhishenhuang/mri}.
\end{abstract}

\section{Introduction}
\label{sec::intro}

Magnetic resonance imaging (MRI) is a widely used imaging technique that allows visualization of both anatomical structures and physiological functions.
%used for scanning images of soft-tissues such as brain or abdomen. 
MRI scanners sequentially collect measurements in the frequency domain (or $k$-space), %and collect raw data 
from which an image is reconstructed. A central challenge in MRI is its time-consuming sequential acquisition process as the scanner needs to densely sample the underlying $k$-space for accurate reconstruction.
%magnetization responses with many different gradient fields. 
In order to improve patients' comfort and safety and alleviate motion artifacts, reconstructing high-quality images from limited measurements is desirable.
There are two core parts in the 
\begin{edits}
\del{conventional}
\end{edits}
accelerated MRI pipeline: a sampling pattern deployed to collect the (e.g., limited/undersampled) data in $k$-space and a corresponding reconstruction method (reconstructor) that also enables recovering any missing information. 
In this work, we use machine-learned models to predict the undersampling pattern and perform image reconstruction in a single 
\begin{edits}
\del{shot or}
\end{edits}
pass and in an object-adaptive manner.

\subsection{Background}

Early approaches to improve the sampling efficiency 
without degrading image quality in MRI include
careful design of the pulse sequence and $k$-space trajectory~\cite{POSER2018,Bitar_06_MRI_review}, partial or half Fourier methods~\cite{Liang_Book}, and hardware-based acceleration techniques such as parallel data acquisition or parallel MRI (p-MRI)~\cite{pMRI-Survey}.
% , which exploit the diversity offered by multiple RF receiver coils and reduce the amount of acquired $k$-space data.
% p-MRI is widely available in commercial scanners.
Despite the availability of several receiver coils, p-MRI is limited to acceleration of typically three to four fold due to increased noise and imperfect artifact correction at higher accelerations. 
% Other early scan acceleration approaches include partial or half Fourier methods~\cite{Liang_Book}.
Recent methods such as compressed sensing (CS)~\cite{emmanuel2004robust}
have leveraged additional structure of priors on the target images to perform reconstruction. MR images 
\begin{edits}
\del{are often structured, e.g., they} 
\end{edits}
can often be approximated as piecewise smooth functions with relatively few sharp edges,
\begin{edits}
\del{. This type of structure}
\end{edits}
which leads to sparsity in the wavelet domain, as piecewise smooth functions are compressible when represented in the wavelet basis. The MR image reconstruction problem from subsampled $k$-space measurements often takes the form of a regularized optimization problem: \begin{equation} 
    %\min_\x \,\frac{1}{2}\|\A \fft\x - \y\|^2 + \alpha\mathcal{R}(\x),
     \min_\x \,\frac{1}{2}\|\A \x - \y\|^2 + \alpha\mathcal{R}(\x),
    \label{reginversion}
\end{equation}
where $\x\in\C^{m\times n}$ is the underlying MR image to recover, 
$\y$ 
%$\y\in \C^{m\times n}$
denotes the partial measurements,
$\A$ is a linear MRI measurement operator,
%$\A$ is a subsampling operator, 
% $\A\in\R^{m\times m}$ is a row-wise subsampling matrix, 
%$\fft$ is the Fourier transform operator,
$\mathcal{R}$ is a regularizer, and $\alpha$ is a regularization parameter that controls the trade-off between the fidelity to the $k$-space measurements and alignment with structure imposed by the regularizer $\mathcal{R}$. 
% We used a single-coil Cartesian measurement operator above, but this could be readily replaced with multi-coil or non-Cartesian forward operators~\cite{pMRI-Survey}.
When representing an MR image as sparse in the wavelet domain, with $\Phi$ as the wavelet transform, and adding the total variation penalty ($\|\cdot\|_{\mathrm{TV}}$),
%in the wavelet domain, with $\Phi$ as the wavelet transform and $\|\cdot\|_{\mathrm{TV}}$ as the total variation, 
the above image reconstruction problem is often formulated as~\cite{lustig2007sparse} \begin{equation}
\label{eqn::classic_img_recon_opt_prob}
\widehat{\x} = \arg  \min_{\x}\,
 \alpha\|\Phi\x\|_1 + \beta \|\x\|_{\mathrm{TV}} + \frac{1}{2}\|\A \x - \y \|_2^2.
%\widehat{\x} = \arg  \min_{\x}\, \alpha\|\Phi\x\|_1 + \beta \|\x\|_{\mathrm{TV}} + \frac{1}{2}\|\A\fft\x - \y \|_2^2.
\end{equation}
In this work, we focus on this sampling-reconstruction problem in the single-coil setting, where $\A := \M \fft$. Here $\fft$ is the Fourier transform operator and $\M$ is a subsampling operator. We discuss how our proposed method readily extends to the multi-coil setting in the methodology section.

\begin{edits}
\del{Compressive sensing \add{(CS)} theory reveals that when the measurement operation \add{applied on a signal} is sufficiently incoherent with respect to the \add{sparsifying transform}, one can exactly recover the %ground truth 
underlying image from significantly fewer %$m$ 
samples \cite{FoucartRauhut_CS_Book}.}
% \add{In CS formulation, the sampling operator $\mathbf{A}$ in \eqref{reginversion} is fixed.} }
\end{edits}

In compressed sensing theory, after fixing a subsampling operator $\mathbf{A}$ which, together with the %domain
sparsifying
transform, has the restricted isometry property (RIP), a non-linear reconstruction method via convex optimization can give an approximation of high quality to the ground truth signal~\cite{CSMRIreview}. Verifying that the RIP holds for a general matrix is an NP hard problem~\cite{Tillmann_NPhard_CS}, and thus early results tend to show that random matrices as subsampling operators satisfy RIP properties with high probability~\cite{Baraniuk_08_RIP_randMats}. A possible $k$-space undersampling scheme in MRI is variable density random sampling~\cite{lustig2007sparse}, where a higher probability is allocated for sampling at lower frequencies than higher frequencies. This sampling scheme accounts for the fact that most of the energy in MR images is concentrated close to the center of $k$-space. In this classic MRI protocol, measurements are collected sequentially using subsampling patterns that are chosen a priori, and an iterative solver such as the proximal gradient descent method~\cite{Huang_PPD} or the alternating direction method of multipliers~\cite{boyd_distributed_2011} is used as the reconstruction method
for the regularized optimization problem. 
Some works have focused on appropriate selection of random sampling patterns for CS MRI~\cite{vasanwala2011,adcock:13:btc,adchanpoonrom13}.

Data-driven approaches can elevate the performance of image reconstruction methods by designing data-specific regularizers $\mathcal{R}$ or by introducing deep learning tools as learned reconstructors. 
For regularizer design, one may replace the $\ell_1$ penalty for the wavelet coefficients and the total variation penalty in~\eqref{eqn::classic_img_recon_opt_prob} with a learning-based sparsity penalty where a sparsifying transform (or dictionary) is learned in an unsupervised fashion based on a limited number of unpaired clean image patches~\cite{xu_12_ldx} or is learned jointly with image reconstruction~\cite{wensailukebres19}. 
\begin{edits}
\del{Alternatively, plug-and-play regularization replaces the proximal step in the typical iterative optimization process for~\eqref{reginversion} by a denoising neural network (or generic denoiser) that essentially functions as an implicit data-based regularizer~\cite{plug_and_play_vbw_13}. 
%For introducing deep learning techniques into building reconstructors, a natural extension of neural-network based regularizer is the unrolled network solver. 
}
\end{edits}
For deep learning based approaches for reconstruction, examples of direct deployment of feed-forward neural networks as denoisers include the U-Net architecture-based networks~\cite{2D_unet} and tandem networks that are trained as generator and discriminator in the GAN framework~\cite{GAN_denoiser_19}. Another approach 
\begin{edits}
\del{related deep learning-inspired approach for reconstruction}
\end{edits} 
involves unrolling iterative algorithms and learning the regularization parameters in a supervised manner from paired training sets. 
\begin{edits}
\del{Each learnable block in the chain of the unrolling strategy is fulfilled by a network to simulate a certain optimization operation~\cite{admm-net2016,NeumannNet_19}.}
\end{edits}
We refer readers to~\cite{Huang_MBIR_review_20,reconreviewsaijeffjong2020,ongie_deep_2020} for detailed reviews of learning-based reconstruction methods. 

\subsection{Contributions}
In this work, we propose a neural-network based adaptive sampler, \mnet{} that generates object-specific sampling patterns in a single-pass based on limited low-frequency information from varying input objects.
\begin{edits}
\del{In this work, we propose an adaptive sampler, \mnet, as a neural network that takes limited low-frequency information as input per scan or frame and outputs corresponding desired undersampling patterns adaptive to different input objects in a single pass. }
\end{edits}
The main advantage of the proposed sampling-pattern predictor \mnet{} is that it outputs adaptive masks with respect to each different input object, and thus the generated sampling pattern varies from case to case. This property gives our sampler more potential compared to several earlier proposed samplers.
\begin{edits}
\del{, as the mask output from our sampler is not only data-driven but also object-specific. }
\end{edits}
Moreover, compared to recent sequential-decision-type samplers~\cite{RL_sampling_20}, our sampling approach determines at once the entirety of the sampling pattern in the high frequency regime, thus reducing the processing time for practical deployment.
While the proposed approach is applied for static MRI sampling in this work, it can be readily used for dynamic MRI sampling (e.g.\ predicting the sampling pattern for a following frame based on low-frequency samples of previous frames).
\begin{edits}
\del{, where variations could also be used to enable faster implementation.}
\end{edits}

The sampling network is trained jointly with a parametric reconstructor to optimize the eventual quality of image reconstruction. We propose an alternating training framework to update the parameters of the sampler network and those of the reconstruction network. The key component in this training framework is the internal label generation mechanism, referred to as \mb{}, which efficiently generates object-specific sampling patterns with respect to each fully sampled $k$-space data point. This training framework makes training an \mnet{} possible without resorting to computationally expensive and approximate greedy methods~\cite{greedy_sampling_mri_18} to supply training labels.

Our numerical experiments show the superior performance of the \mnet{} framework with respect to several alternative schemes for undersampled single-coil acquisitions based on the public FastMRI~\cite{fastMRI_dataset_18,fastMRI_journal} dataset.
\begin{edits}
\del{In this work, we develop an adaptive sampler that generates object-specific sampling patterns based on highly limited measurements from varying input objects.
We propose to leverage neural networks to generate undersampling patterns in a single pass using highly undersampled initial acquisitions as input. The sampling network is trained jointly with a reconstructor to optimize eventual image reconstruction quality. Such training of neural networks to predict sampling patterns is essentially a supervised task. 
%In order to obtain training labels  in an efficient manner, 
We propose an end-to-end training procedure that takes in fully sampled $k$-space data 
% (and corresponding reconstructions) and initial acquired data 
and outputs corresponding desired undersampling patterns and image reconstructions. To ignite the process of training a network for sampling pattern or mask prediction, we use an alternating framework that operates between finding object-adaptive binary masks (and a corresponding reconstructor) %with respect to 
exploiting ground truth $k$-space data, 
and updating a network for predicting such sampling masks. 
%(that serve as internally generated labels). 
%The sampler and reconstructor are trained jointly in the training pipeline in an end-to-end fashion.
}
\end{edits}

\subsection{Connection to Bilevel and Mixed-integer Optimization Problems}
The drawback of the conventional compressive sensing approaches~\cite{lustig2007sparse,adchanpoonrom13} to accelerate MRI scans
%of MR image reconstruction problem 
is that the measurement operator does not adapt to different objects and thus may not result in optimal image reconstructions in general. 
%given certain constraint relaxation or iterative solvers.

An alternative formulation, first explored in~\cite{saiembc2011},
is in the form of a bilevel optimization problem:
\begin{align}
\label{prob::bilevel_opt}
    &\min_{\boldsymbol{\theta}} \sum_{i=1}^N \frac{1}{2}\|\widehat{\x}_i(\boldsymbol{\theta}) - \x_i^{\textrm{gt}}\|_{2}^{2} \nonumber\\
    &\textrm{s.t. } \widehat{\x}_i( \boldsymbol{\theta} ) = \mathrm{arg}\,\min_\x \frac{1}{2} \|\y_i^{\boldsymbol{\theta}} -\A_{\boldsymbol{\theta}}\x\|_2^2 + \lambda \mathcal{R}(\x) \quad \forall \,i,
\end{align}
where $\boldsymbol{\theta}$ denotes the underlying parameters of the measurement/sensing operator $\A_{\boldsymbol{\theta}}$; $\x_i^{\mathrm{gt}}$ are the ground truth images reconstructed from raw fully-sampled $k$-space data $\y_i^{\mathrm{gt}}$; $\y_i^{\boldsymbol{\theta}}$ are the corresponding measurements from applying the measurement operator $\A_{\boldsymbol{\theta}}$; and $\mathcal{R}(\x)$ is the regularization term for the lower level optimization problem. The goal is to ultimately generate %noise-free and aliasing-free 
artifact-free
images from partial measurement vectors $\{\y_i\}$
%observation $\{\widehat{\x}_i\}$, 
%which is 
as embodied in the upper level optimization problem. This particular formulation accommodates variation in the measurement operator, thus rendering the operator to be adaptive to a training dataset. 
%or potentially to specific objects (if the 
%object adaptive. 
In the MRI setting, the parameters $\boldsymbol{\theta}$ capture the sampling patterns (\textit{masks}) we want neural networks to predict. A major challenge in tackling the bilevel optimization formulation is the \textit{implicit} dependence of the upper-level objective on the parameters $\boldsymbol{\theta}$. It is not clear in either theoretical or practical aspects how to compute the (sub)gradients of the upper-level objective with respect to the pattern parameters. To address this issue, we use a neural network as the solver for the lower-level problem, which essentially renders the `$\textrm{arg} \min$' in formulation \eqref{prob::bilevel_opt} as $\widehat{\x}_i(\boldsymbol{\theta}) = \textrm{Network}(\y_i^{\boldsymbol{\theta}})$, and therefore the gradient computation and back-propagation from the upper-level loss function to the sampling parameters $\boldsymbol{\theta}$ become possible.

In this work, we focus on Cartesian undersampling of $k$-space.
%the sampling pattern we consider are line sampling patterns in the $k$-space. 
In this case, we can write $\A_{\boldsymbol{\theta}} =\mathrm{diag}(\M)\fft$, where $\mathrm{diag}(\M)$ is a binary diagonal matrix, and a $1$ in a certain entry of $\M$ indicates the corresponding row in $k$-space is observed. Namely, the parameters $\boldsymbol{\theta}$ in the bilevel formulation \eqref{prob::bilevel_opt} are the binary elements in $\M$. A practitioner can pre-set a sampling budget $b$, which is reflected as $\sum_t M_{t} = b$. With a sampling budget constraint, the optimization problem \eqref{prob::bilevel_opt} assumes the flavor of integer programming:
\begin{align}
\label{prob::milp}
    &\min_{\M} \sum_{i=1}^N \frac{1}{2}\|\widehat{\x}_i(\M) - \x_i^{\textrm{gt}}\|_{2}^{2} \nonumber\\
    &\textrm{s.t. } 
    \begin{dcases}
    \widehat{\x}_i( \M ) = \mathrm{arg}\,\min_\x \frac{1}{2} \|\y_i^{\M}-\mathrm{diag}(\M)\fft\x\|_2^2 + \lambda \mathcal{R}(\x) \quad\forall \,i,&\\
    \sum_{t=1}^m M_{t}= b & \\
    M_{t}\in\{0,1\} \quad \forall \,t\in[m].&
    \end{dcases}
\end{align}
There has been work devoted to leveraging neural networks to construct solvers for mixed-integer programming (MIP). In~\cite{NN_for_MIP_20}, a mixed-integer program is represented as a bipartite graph (nodes are variables and constraints of the MIP, and edges connect variables and their corresponding constraints). One neural network is trained to predict integer values in the solution to cater for the integer constraints in a generative style, and another one is trained to learn the policy of branching. Classic solvers are used to tackle sub-problems of the original MIP to provide training data for each network.

The target of predicting a sampling pattern echoes the integer-valued constraint in the optimization problem~\eqref{prob::milp}. We address the integer-valued constraint in our approach through a binarization step, which can be considered as a post-processing action applied on the output of the proposed mask-predicting \mnet{}. While the way we deploy the network for mask prediction does not constitute directly a solver for the MIP, it gives a potential new direction of using networks for integer-constraint problems.

\subsection{Prior Art in Sampling Design}
We now briefly review several existing sampling design strategies in the literature. 
%which differ fundamentally in their design ideas. 
An early work optimizes $k$-space trajectories using an information gain criterion~\cite{seeger2010}. Another first work~\cite{saiembc2011} directly optimizes a $k$-space undersampling pattern based on reconstruction error-type loss on a training set.
% , and employs an approximate algorithm for training.
With recent success of deep learning methods, there has been growing interest in the topic of adaptive sampling. A classical non-parametric sampling adaptation method for~\eqref{prob::milp} is the greedy approach~\cite{greedy_sampling_mri_18}, which chooses one $k$-space line or phase encode
%high frequency 
to add into the sampled set at each step depending on which line in combination with those already added so far leads to the best reconstruction loss (upper level loss). The greedy approach terminates when the sampling budget is reached. The learned sampling mask is population-adaptive (learned on a dataset) but does not vary with different objects (i.e., not object-adaptive).
%by comparing the quality of reconstructed images from observed low-frequency information plus one high frequency candidate~\cite{greedy_sampling_mri_18}. 
%The greedy sampling process is independent from the reconstructor used in the practice, and 
The greedy sampling process 
can be accelerated by considering random subsets of all candidate high frequencies as cohort and selecting multiple high frequencies into the sampling set at one step. 

LOUPE is a parametric model that characterizes the probability of sampling each pixel or row/column in the frequency domain or $k$-space with underlying parameters that are learned simultaneously with the parameters of a reconstructor~\cite{Loupe_20}. The drawback of this approach is that the learned sampling pattern is generic with respect to the training set (population-adaptive) rather than being adaptive to each individual object's characteristics. A similar work~\cite{Sherry_sampling_mri_20} also assumes that the training data are representative enough for new data acquisitions and learns a fixed undersampling pattern.
%hence outputting a fixed sampling pattern as the learnt result. 
This latter approach combines the sampling and image reconstruction problem as a bilevel optimization problem, parametrizes the sampling pattern via probability variables to optimize, and resorts to iterative optimization solvers to find the solution. 

J(oint)-MoDL~\cite{JMoDL_20} assumes the sampling pattern to be rows/columns sampled in $k$-space and
learns a common undersampling pattern or mask along with a neural network reconstructor from a training set. 
%the parameters to tune determine which rows/columns to sample \cite{JMoDL_20}. 
Another way of parametrizing sampling patterns is to assign each frequency a logit as the natural logarithm of the unnormalized class probability so that a categorical distribution with respect to all frequency candidates can be accordingly constructed by applying a softmax function on the associated logits, which enables gradient back-propagation, and the output mask contains the frequencies with top-M weights~\cite{generative_samp_20}. PILOT parametrizes the sampling patterns as a matrix to coordinate with the non-uniform FFT measurements, and adds in an additional round of optimization to enforce hardware measurement constraints~\cite{Pilot_mri_2019}. 
The BJORK method parametrizes sampling patterns as linear combinations of quadratic B-spline kernels, and attempts to reduce the number of parameters needed to characterize sampling patterns through a scheduling plan that gradually reduces the ratio between the total number of $k$-space samples and the number of B-spline interpolation kernels~\cite{BJORK_21}.  
The approach can handle non-Cartesian sampling but does not learn an object-adaptive sampler.

Sequential decision processes have also been exploited as a path to build samplers. One attempt in the reinforcement learning style formulates the sampling problem as a partially observable Markov decision process and uses policy gradient and double deep Q-network to build the sampling policy~\cite{RL_sampling_20}. Another attempt builds a neural network as the sampler to be repeatedly applied in the sampling process and trained simultaneously with the reconstructor~\cite{Bouman_samp_21}. 
\begin{edits}
\del{The sequential methods consume additional time for each prediction/decision (idle scanner time) that can introduce artifacts in real-time imaging.}
\end{edits}

%The time lost for making sequential prediction/decision could introduce artifacts, especially in dynamic
%when used in real-time imaging.

\subsection{Structure of this Paper}
The rest of this paper is organized as follows. Section~\ref{sec::method} elaborates the design of the \mb{} training framework. Section~\ref{sec::experiments} discusses the architecture of \mnet{}, the implementation details and experimental results. We summarize our findings and potential new directions for future work in section \ref{sec::conclusions}.

\section{Methodology} 
\label{sec::method}
MRI scanners acquire measurements in the frequency domain ($k$-space). We let $\x\in\C^{m\times n}$ denote an underlying (ground truth) image  and $\fft\x\in\C^{m\times n}$ denotes its Fourier transform. 
Here, we focus on Cartesian (or line) sampling of $k$-space and single-coil data in the experiments of this work, 
%We consider the 1D line sampling in this work, and thus we assume the sampling pattern 
where we represent the sampling mask parameters (when doing subsampling of rows in $k$-space) via the vector $\M$ in $\R^m$. 
In this case, the fully-sampled $k$-space (including noise) is related to the corresponding underlying image as $\fft\x$.
%$\M$ is a vector in $\R^m$. 
The subsampling of $k$-space with zeros at non-sampled locations 
can be represented as 
$\widehat{\y} = \mathrm{diag}(\M)\fft[\x]$. 
%$\widehat{\y} = \mathrm{diag}(\M)\y = \mathrm{diag}(\M)\fft[\x]$. 
More generally, for multi-coil data, 
%the subsampling is performed for each coil's fully sampled $k$-space
the coil sensitivity maps would scale the image $\x$ in this process~\cite{pMRI-Survey}.
We define the acceleration factor as the ratio between the number of all available rows (alternatively columns) in $k$-space and the total number of subsampled rows (i.e., $\sum_i M_i$). 

\subsection{Why do We Need the \mb{} Routine?}
We aim to train a neural network that can generate a subsampling pattern for $k$-space based on limited initial low-frequency measurements. An immediate ensuing challenge for training such a network in the supervised style is the lack of labels denoting ideal/best subsampling patterns. 
One plausible option is
%may 
to generate object-specific masks
for each training image using the greedy algorithm~\cite{greedy_sampling_mri_18} for~\eqref{prob::milp}
and use them as labels. However, the greedy algorithm's labels are not necessarily optimal (i.e., global minimizers in~\eqref{prob::milp}) and
obtaining such labels for a large number of training images (from fully-sampled $k$-space) is computationally expensive or infeasible.

Another natural strategy to circumvent the difficulty of obtaining adaptive mask labels is an end-to-end pipeline that is comprised of an encoder, which maps limited (e.g., low-frequency) information in $k$-space to intermediate adaptive masks, and a decoder that attempts to reconstruct the artifact-free image from the $k$-space information sampled according to the adaptive masks from the encoder. After training of the end-to-end pipeline is concluded, one extracts the encoder part from the pipeline, which can function as a desired mask prediction network. We choose not to proceed with this obvious end-to-end approach owing to two folds of consideration. 
The first concern is that binarizing intermediate masks to meet sampling budget target of the mask prediction network is incompatible with gradient-based methods.
% The mask-predictor \mnet{} should have binary labels in the training process. 
While the computation techniques can still get around the hard binarization thresholding in gradient computation and back-propagation, there are not binary labels in the end-to-end training framework to supervise the mask prediction network, which may incur unexpected behavior in its performance. 
% , and moreover there are no binary labels for guiding the mask prediction network in the end-to-end training framework. %, as the only labels are the ground truth images. # Leo's edit: Nov 16
%The first concern is that the end-to-end approach cannot enforce exact binary masks or exact sampling budget on the \mnet{} output (or involve some sort of binary labels) as this would be problematic for gradient computation with respect to \mnet{} parameters for updating them.
The second concern is that a deep end-to-end training procedure without intermediate intervention can cause vanishing gradients, which results in slow progress in training. In addition, an end-to-end approach makes it harder to track the performance of each component and thus has less interpretability for the outcome of each component in this framework.
We alleviate these issues by introducing a split variable that represents the output of \mnet{} and 
satisfies the explicit mask constraints in~\eqref{prob::milp}. A penalty is then introduced to measure the fidelity of the \mnet{} output to the split binary variables/labels.

\subsection{Proposed Training Scheme}
\begin{figure*}[!t]
  \centering
  \includegraphics[width=\textwidth]{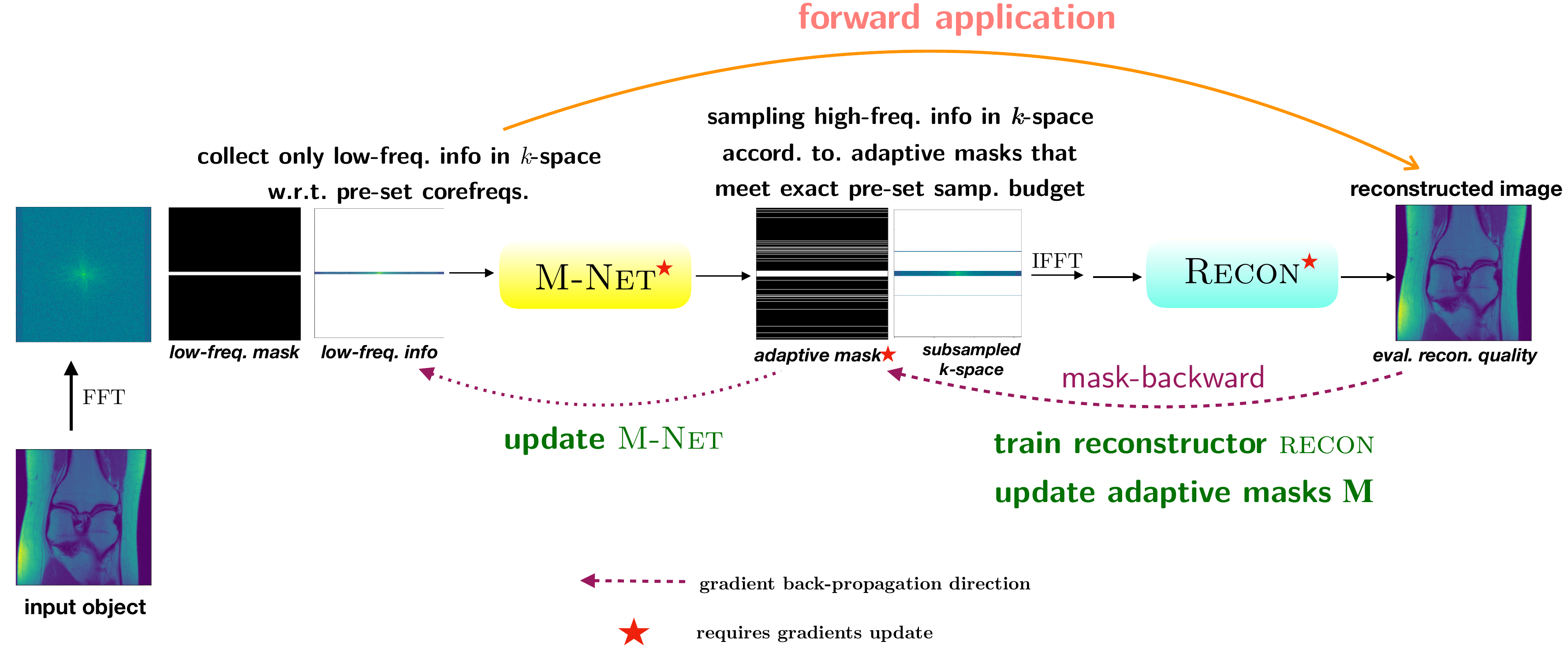}
  \caption{Diagram of the proposed alternating training framework.}
  \label{fig::pipline}
\end{figure*}
We propose a \mb{} inferring method (see Subroutine~\ref{alg::mb}) to obtain binary label masks to feed the supervised training process of the mask-predicting network (\mnet). The key component of the \mb{} method is a parametric reconstructor that serves as a bridge connecting mask parameters and ground truth images. 
With the mask $\M$ as a parameter to tune, we first compute a zero-filled image reconstruction via inverse Fourier transform (\texttt{IFFT}) of the observed low-frequency $k$-space data with zeros filling the unobserved $k$-space. 
The reconstructor receives this crude initial reconstruction and refines it to be free of aliasing artifacts. 
The loss function in~\eqref{eqn::mb_loss} compares the refined image with the ground truth image using specific reconstruction quality metrics,
and is optimized with respect to object-adaptive sampling masks and the reconstructor.
%and then back-propagating the gradients to refine sampling mask $\M$. 
We mainly use a standard 8-block U-Net as the reconstructor with additional residual connections between the downsampling and upsampling blocks. Later, we consider other reconstructors in combination with $\mnet$. 
%During the gradient back-propagation, the parameters of the Unet reconstructor are also updated simultaneously with $\M$.

\begin{algorithm}[!t]
\caption{Subroutine: \mb{}}
\label{alg::mb}
\begin{algorithmic}[1]
\Require Ground truth image $\x^\star$, an initial mask $\M_\textrm{init}$, a reconstructor $U$ with parameters $\Theta$, maximal iteration steps $T$, sparsity control parameter $\alpha$, consistency control parameter $\lambda$, sampling budget $b$ % \Comment{with slight overburderning of notation, $\x^\star$ can mean a single image or a batch of images}
\State Initialize $\M(\boldsymbol{\xi}_0) = \M_\textrm{init}$
\For{iteration count $t<T$}
    \State Sample information in $\fft[\x^\star]$ according to binarized $\M(\boldsymbol{\xi}_t)$ to obtain adaptive observations $\y_t$.
    \State Compute the zero-filled \texttt{IFFT} images $\x^{\texttt{IFFT}}_t =\fft^{-1}[\y_t]$. 
    \State Compute the refined image $\x^{\textrm{recon}}_t = U_{\Theta_t}(\x^{\texttt{IFFT}}_t)$.
    \State Evaluate the loss function in~\eqref{eqn::mb_loss}, compute the gradient with respect to $\boldsymbol{\xi}_t$ and parameters $\Theta$ of the reconstructor $U$, make a step of update to obtain $\boldsymbol{\xi}_{t+1}$ and $\Theta_{t+1}$. \Comment{See Remark 2 for implementation details.}

\EndFor
\State Apply sampling budget control $b$ and binarize $\M(\boldsymbol{\xi}_T)$ to obtain $\M_\textrm{out}$.
\State \textbf{return} $\M_\textrm{out}$, image reconstructor $U_{\Theta_T}$ \Comment{This subroutine can readily handle a batch of images.}

\end{algorithmic}
\end{algorithm}

In order to train the \mnet{} with guidance from the \mb{} inferring process, we use the following alternating training framework. Let $\z_i$ denote the initially observed low-frequency information of the ground truth image $\x_i^\star$, $\Gamma$ denotes the parametrization of the \mnet{}, and $\Theta$ denotes the parametrization of the reconstructor \texttt{Recon}. The optimization problems involved in this training framework are the following that are solved in an alternating manner (Fig.~\ref{fig::pipline}): \begin{align}
&\mb:\nonumber\\
&\begin{aligned}
    \label{eqn::mb_loss}
    \min_{\{\M_{\textrm{b},i}\},\Theta}\, \sum_{i} &\bigg( \underbrace{\frac{\| \texttt{Recon}_{\Theta}\big(\fft^{-1} \mathrm{diag} (\M_{\textrm{b},i}) \fft\x_i^\star \big) - \x_i^\star \|_2}{\| \x_i^\star \|_2}}_{\textrm{reconstruction data fidelity }}  \,\, +  \, \alpha \underbrace{\mathcal{R}\big( \M_{\textrm{b},i} \big)}_{\textrm{sparsity reg.\ }} + \,\,
    \lambda \underbrace{\varphi\big(\mnet_\Gamma (\z_i) , \M_{\textrm{b},i} \big)}_{\textrm{consistency}} \bigg),
\end{aligned}\\
&\mnet:\nonumber\\
&\begin{aligned}
\label{eqn::mnet_loss}
    &\min_{\Gamma} \sum_{i} \varphi\big(\mnet_{\Gamma}(\z_i), \M_{\textrm{b},i} \big).
\end{aligned}
\end{align}
In loss function~\eqref{eqn::mb_loss} of \mb{}, the first term characterizes the fidelity between the reconstructed image and the ground truth image using normalized root mean square error (NRMSE), the second term is the regularization term to implicitly control the sparsity of the returned mask, and the third term imposes the consistency between the returned mask and the mask predicted by the current \mnet{}. We include the consistency term to accelerate the convergence of this alternating training framework. We take $\mathcal{R}(\M_{b,i}) = \|\M_{b,i}\|_1$ to promote sparsity on the object-adaptive masks.
The exact binary constraint and desired sampling budget (as in~\eqref{prob::milp}) are enforced on $\M_{b,i}$ after the gradient step as shown in Subroutine~\ref{alg::mb}.
The function $\varphi$ denotes the binary cross entropy loss in both loss functions~\eqref{eqn::mb_loss} and~\eqref{eqn::mnet_loss}. The non-negative parameters $\alpha$ and $\lambda$ 
%are magnitude hyperparameters for 
are weights for the sparsity regularization and consistency regularization terms, respectively.

In this training process, we separate the training of \mnet{} and the image reconstructor \texttt{Recon} into two alternating steps to enable easier training. We first use the \mb{} method to find the ideal mask $\M_{\textrm{b},i}$ for each given image $\x_i$ along with updating the reconstructor. 
We start our alternating training framework with a warmed-up reconstructor (e.g., \texttt{Unet}) $U_0$ that is pre-trained for de-aliasing zero-filled \texttt{IFFT} reconstructions from variable density random undersampling of $k$-space.
% the \texttt{IFFT} images from random mask filtered zero-filled observation in the $k$-space.}
% with its inputs (initial reconstructions) set to zero-filled reconstructions obtained with variable density random undersampling of $k$-space.}

Generating mask labels through \mb{} method and updating the mask predicting network are executed alternatively. %until convergence. 
%In summary, we have the following 
The complete training algorithm is summarized in
Algorithm~\ref{alg::alternating_training}.
    
Since the \mb{} process involves nontrivial gradient update and post binarization, prior to updating \mnet{},
we use the reconstruction data-fidelity as a quality function,  $\mathcal{Q}(\M,U,\x^\star)$, to evaluate how good a mask $\M$ is with the reconstructor $U$ for the image $\x^\star$. We define $\mathcal{Q}(\M,U,\x^\star) = -\frac{\|U(\fft^{-1}\mathrm{diag}(\M)\fft\x^\star) - \x^\star\|_2}{\|\x^\star\|_2}$.  
We need to ensure that the refined mask is improved in terms of contributing more for image reconstruction and that the refined mask is non-trivial. Therefore, the condition to accept a refined mask is that its quality should be better than that of the previous direct output from \mnet{} (pre-refinement mask) and that of random masks. The quality of the random mask is always evaluated with the initial reconstructor $U_0$ pre-trained for random-masked observations. Meanwhile, the quality of adaptive masks and their refined counterparts is evaluated by the reconstructor $U_s$ that is updated during training.
    
\begin{figure*}[!ht]
  \centering
  \includegraphics[width=\textwidth]{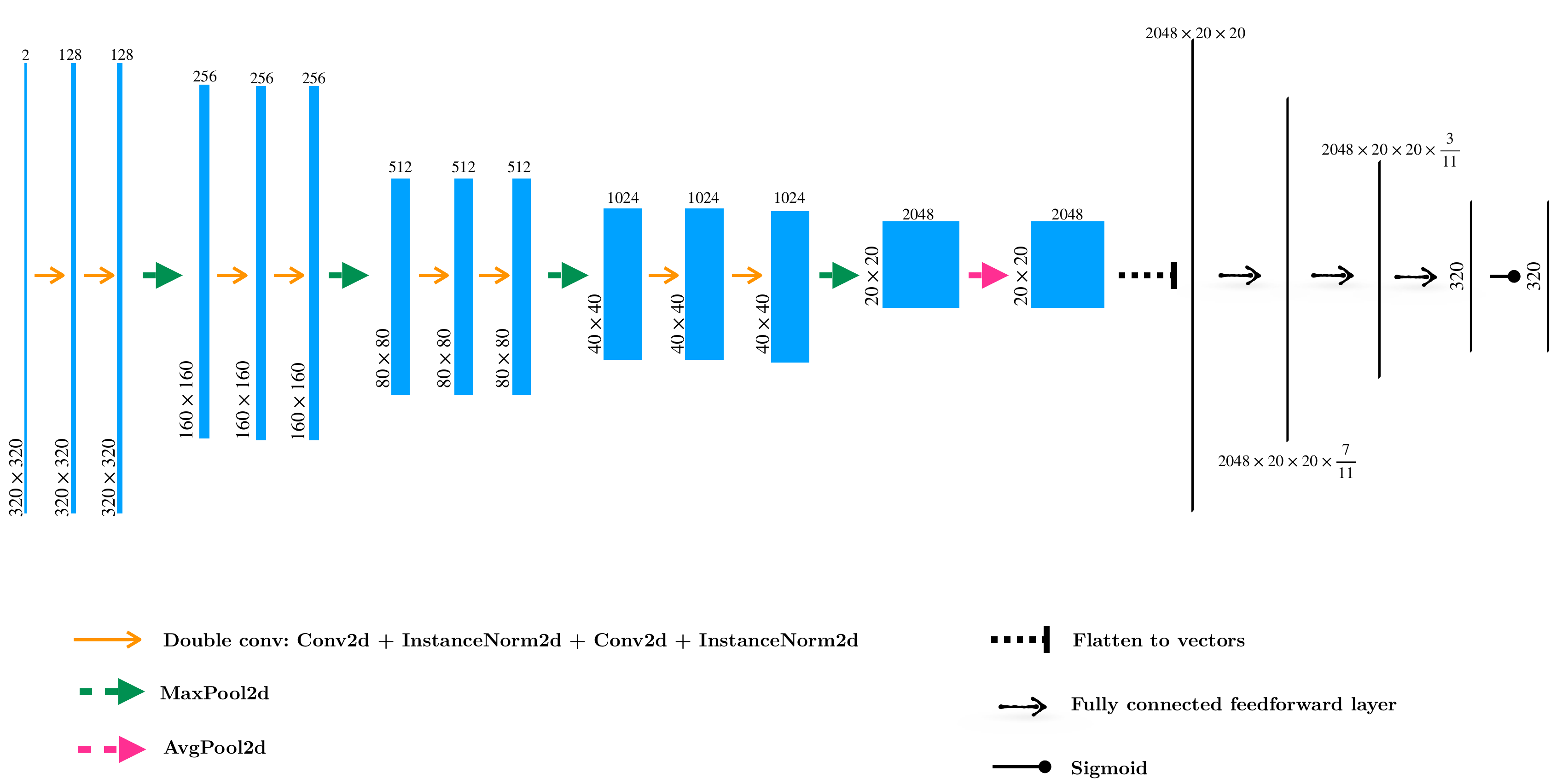}
  \caption{Structure of \mnet{} used in training. The kernel of 2D convolution is of size $3\times 3$, stride and dilation of sizes $1\times 1$ and no padding. The Double-Conv layers maintain the size of the image for its input and output. The kernel size of the MaxPool layer is $3\times3$, and the kernel size of the AvgPool layer is $2\times 2$.}
  \label{fig::mnet_structure}
\end{figure*}

After a refined mask passes the quality check, we use the ideal binary $\M_{\textrm{b},i}$ to update the parametrization of the \mnet. The direct output of the deployed \mnet{}, as shown in Figure \ref{fig::mnet_structure} represents the significance of each high frequency. Before we evaluate the loss function $\varphi$ in equation \eqref{eqn::mnet_loss}, we need to process the primitive output \mnet{$_\Gamma(\z_i)$} to obtain a binary mask, and ensure that this output mask has the target sampling ratio. 

Regarding \textit{sampling budget control}, we have used the normalization trick in~\cite{Loupe_20} to ensure the output mask from \mnet{} as well as the refined mask returned by the \mb{} method has the exact target sampling ratio. Assume the sampling budget of lines is $b$. Thus, the sampling ratio is $\alpha=\frac{b}{m}$, or for a corresponding unbinarized mask $\M$, it is $\frac{\|\M\|_1}{m}=\frac{\|\sigma(\boldsymbol{\xi})\|_1}{m}$, where $\sigma(\cdot)$ is the sigmoid function and $\boldsymbol{\xi}$ is mask parametrization parameters (see Remark 2). Assume $\mathbf{P}$ is the mask characterization before normalization, thus $\|\mathbf{P}\|_1\ne b$. We define $\widetilde{p} = \frac{\|\mathbf{P}\|_1}{m}$, which is the pre-normalization sampling ratio, and thus $1-\widetilde{p}$ is the average value of $1-\mathbf{P}$. The normalization is done by
\begin{equation}
        N_\alpha(\mathbf{P}) = 
        \begin{cases}
        \frac{\alpha}{\widetilde{p}}\mathbf{P}, \, \textrm{if } \widetilde{p}\ge \alpha\\
        1-\frac{1-\alpha}{1-\widetilde{p}}(1-\mathbf{P}), \, \textrm{otherwise}.
        \end{cases}
\end{equation}

The \textit{binarization} is necessary to make ultimate output sampling masks meaningful for practical use. The threshold we use for binarizing vectors is 0.5. During the training process, each entry in the mask characterization 
$\M$ is in $(0,1)$. We should point out that the reconstructor in 
Subroutine~\ref{alg::mb} %(in our experiments, 
(e.g., the Unet or the MoDL~\cite{MoDL_18} model) sees the subsampled $k$-space information $\y_t$ filtered by the binarized mask, while the gradient with respect to unbinarized $\M$ in~\eqref{eqn::mb_loss} can still be computed with the definition of a (approximate) \textit{backward} function in e.g., the PyTorch implementation of binarization (see~\cite{Bengio_13_gradient,Loupe_extension_20,Bouman_samp_21}).
    
The sampling budget control and binarization are both applied before a refined mask is returned by the \mb{} method to update the \mnet{} model, thus rendering the labels for training the \mnet{} to have the pre-specified sampling ratio. They are also applied on the direct output of \mnet{} at testing time.

\begin{algorithm}[!ht]
\caption{Alternating training framework for sampler network \mnet{} and reconstructor}
\label{alg::alternating_training}
\begin{algorithmic}[1]
\Require A mask-predicting network \mnet{}, a reconstructor $U_0$, ground truth images $\{\x^\star_s\}$, initial (low-frequency) measurement count $l$, sampling budget $b$, maximal iteration steps $T$ for \mb, sparsity control parameter $\alpha$, consistency control parameter $\lambda$, number of iterations $c$ for \mnet{} updates.

\For{epoch $<$ maximal epoch}
    \For{$s < \textrm{maximal batch number}$}
        \State Compute $\y_s = \fft[\x^\star_s]$ and extract $k$-space information $\z_s = \y^{\textrm{lf}}_s$ corresponding to lowest $l$ frequencies.
        \State Generate the adaptive masks $\M_s = \mnet{}(\z_s)$ for image batch $\x^\star_s$.
        \State Apply sampling budget control with budget $b$ and binarization on $\M_s$. % \Comment{See remarks 3 and 4.}
        \State $\allowbreak [\M^{\textrm{refined}}_s, \widetilde{U}_s] = \mb(\x^\star_s,\M_{s}, U_s, \allowbreak T,\alpha,\lambda, b)$
        \If{$\mathcal{Q}(\M^{\textrm{refined}}_s,\widetilde{U}_s,\x^\star_s) > \max\big\{ \mathcal{Q}(\M_s,\widetilde{U}_s,\x^\star_s), \allowbreak \mathcal{Q}(\M_{\textrm{random}},U_0,\x^\star_s)\big\}$}
        % \If{quality of $\M^{\textrm{refined}}_k$ $> \max($quality of $\M_{s}$, quality of $\M_{\textrm{random}})$}
            \State Make $c$ steps of supervised training of \mnet{} with labels as the refined mask $\M^{\textrm{refined}}_s$ with respect to objective \eqref{eqn::mnet_loss}.
            \State $U_{s+1} = \widetilde{U}_s$
        \Else
            \State $U_{s+1} = U_s$
        \EndIf
    \EndFor
\EndFor
\State \textbf{return} \mnet, image reconstructor $U$

\end{algorithmic}
\end{algorithm}

Next, we list a set of remarks about our approach and some extensions.

\subsection{Remarks} \begin{enumerate}
    \item Algorithm~\ref{alg::alternating_training} follows a co-design approach to identify a sampler and a reconstructor simultaneously with respect to a training dataset. Finally, one can also take a trained \mnet{} sampler from the output of Algorithm~\ref{alg::alternating_training} and train a %corresponding 
    reconstructor in a separate process. We have investigated the performance of both of these practices in Section~\ref{sec::experiments}.
    
    \item \textit{Computing gradient with respect to mask parametrization parameter} $\boldsymbol{\xi}$: When implementing~\eqref{eqn::mb_loss}, we let $\M_{b,i} = \sigma(\boldsymbol{\xi}_i)$ where $\sigma(\cdot)$ is the typical sigmoid function and $\boldsymbol{\xi}_i$ is the parameter vector in $\R^m$ characterizing the mask for image $\x_i$. Compared to parametrizing the adaptive mask corresponding to the image $\x_i$ using nonnegative real numbers between 0 and 1 (ideally binary), using sigmoid improves the numerical stability when updates are made to the free $\boldsymbol{\xi}_i$'s.
    
    \item For the mask initialization step in Subroutine~\ref{alg::mb}, the input mask $\M_\textrm{init}$ is binarized, and we set $\boldsymbol{\xi}_0[i] = 0.1$ for frequencies with mask value 1 and $\boldsymbol{\xi}_0[i] = -0.1$ for frequencies with mask value 0. %, which worked well.
    
    \item One can replace a Unet with other parametric reconstructors such as MoDL and unrolling iterative blocks in the \mb{} method in our joint training framework, as long as the reconstructor is differentiable to let the gradient to propagate through.
    
    \item In Step 4 in Algorithm~\ref{alg::alternating_training}, if $\M_s$ are repetitive for more than half of the images in a batch, then we use random masks as the input for the \mb{} process to avoid any degeneracies.
    
    \item \textit{Multi-coil setting}: In this case, the $k$-space measurements of an image are obtained with $n$ coils. For each coil, the measurements can be expressed as multiplication of the image with coil sensitivity maps, followed by the Fourier transform and subsampling of $k$-space (same across coils).
    %can be expressed as $\mathcal{A}:=\sum_{i=1}^{n} \A\circ\fft\circ\mathbf{S}_i$, where $\mathbf{S}_i$ is the sensing operator for $i$-th coil. 
    To apply the \mb{} procedure in the multi-coil setting, we only need to replace the \texttt{IFFT} operation in step 4 of subroutine~\ref{alg::mb} with the adjoint of the multi-coil measurement operator $\A$. 
    % The reconstructor will have to have 2 input channels and 2 output channels to accommodate the real and imaginary parts respectively.
    The input to \mnet{} can have multiple channels to accommodate data from different coils and a learnable convolution across coils/channels can be added as a first step in the \mnet{} architecture to explore correlation among coils. After this first step, the architecture can stay the same as in the single-coil case.
   
   % The structure of \mnet{} can also stay the same when one assembles frequency data from all coils. 
    
\end{enumerate}

\section{Experiments}
\label{sec::experiments}
Here, we first discuss the network architectures in our framework along with datasets used, hyperparameter choices, and the general experiments, before presenting results and comparisons.

\subsection{Implementation Details}
\paragraph{\mnet{} structure}
The \mnet{} mask predictor starts with a double-convolution block that is followed by four downsampling blocks, which are used in the classic Unet, and ends with four fully-connected layers. The input to the \mnet{} is the initially collected appropriately zero-padded limited (low-frequency) $k$-space information with real and imaginary parts separated into two channels, and the output is the sampling decision with respect to unobserved (higher) 
frequencies. The design of \mnet{} consists of two parts: the encoding convolutional layers that emulate the encoder part in the Unet structure, and fully connected feedforward layers which originated in classification networks.

\paragraph{Reconstructors} For Unet reconstructors, we adopt a standard 8-block Unet architecture following~\cite{2D_unet}. When carrying out the \mb{} training, both the input to and output from the reconstructor are single-channel real-valued images (to correspond with magnitude of reconstructions).
For the input, we take the magnitude of the inverse FFT initial reconstruction.
In the separate training process for individual reconstructors, the input images have 2 channels (for a complex-valued image, the real and imaginary parts are separated into 2 channels), which led to slightly better performance,
and the outputs are single-channel real-valued images. The Unet reconstructor contains four downsampling blocks and four upsampling blocks, each consisting of two $3\times3$ convolutions separated by ReLU and instance normalization. We let the first downsampling block to have 64 channels which are expanded from the input 1 or 2 channels. The Unet used in the \mb{} process has the residual structure with skip connections, while the Unet used in the separate training cases does not.

We also briefly review the idea of the MoDL reconstructor~\cite{JMoDL_20}. The MoDL reconstructor consists of a neural-network denoiser and a data-consistency-enforcing block. The image reconstruction problem the MoDL reconstructor tackles is of the form $\widehat{\x} = \mathrm{arg}\,\min_{\x} \|\y - \A\x\|_\mathrm{F}^2 + \|\x - \mathcal{D}_{\Phi}(\x)\|_\mathrm{F}^2$, where $\mathcal{D}_\Phi(\cdot)$ is a neural-network denoiser. The two successive or alternating steps in implementing the MoDL reconstructor are specifically $\x_{n+1} = (\A^\mathrm{H}\A + \mathbf{I})^{-1} (\mathbf{z}_n + \A^\mathrm{H}\y)$ (the data-consistency-enforcing block) and $\mathbf{z}_{n+1}=\mathcal{D}_{\Phi}(\x_{n+1})$ (the denoising step), where $\A^\mathrm{H}$ is the Hermitian of operator $\A$. One typically uses the conjugate gradient (CG) algorithm to compute the inverse in the data-consistency-enforcing block. In the implementation of MoDL in this work, we set the MoDL reconstructor to have 4 MoDL blocks owing to the consideration of computation complexity in the hardware and the error tolerance of the CG algorithm is set to be $5\times 10^{-5}$. The neural-network denoiser used in the MoDL reconstructor is a Unet as previously described with 64 channels in the first downsampling block without residual structure. 

\paragraph{Dataset} We used the single-coil data in the NYU FastMRI dataset~\cite{fastMRI_dataset_18,fastMRI_journal} for training and testing. In each file, we selected the middle 6 slices containing most prominent image patterns
in the knee scan, thus making up a training dataset with 12649 images, and 1287 images were used for validation and 1300 images were used for testing (we follow the setup of~\cite{RL_sampling_20} and split the original fastMRI validation set into a new validation set and a test set with the same amount of volumes). Each slice has the dimension $320\times 320$, and so does its corresponding $k$-space. %slice.

\paragraph{Baselines} With the same sampling ratio and low-frequency base (initial) observations, We benchmark our adaptive sampling \mnet{} framework against the recent LOUPE sampler + reconstructor scheme, a random sampler, an equi-distance sampler, and an energy-density based sampler~\cite{energy_sampling}. For the energy-density based sampler, we constructed the sampling probability distribution as follows. With respect to each image, we compute the energy of each row of its $k$-space and normalize the energy vector for that image. We then take the average of all normalized energy vectors for each image in the training dataset as the energy-based sampling probability distribution. We observed that the sampling probabilities from the energy-based density heavily concentrate in the low-frequency regime.
We compare the performance of different methods for 4$\times$ and 8$\times$ acceleration of $k$-space sampling, respectively. For a fair comparison to the random and equi-distance samplers, corresponding reconstructors are trained for both these cases.

\paragraph{Hyper-parameters} We used the normalized $\ell_2$ loss and the structural similarity index measure (SSIM) for measuring the image reconstruction quality. SSIM is computed using a window size of $11\times 11$ and hyperparameters $k_1=0.01$, $k_2=0.03$.  When training a reconstructor (Unet or MoDL) in a separate manner (with respect to inputs sampled with \mnet{} masks, random masks and equi-distance masks), the loss function is set as $\frac{\| \x_{\textrm{recon}} - \x_{\textrm{gt}}\|_2}{\|\x_{\textrm{gt}}\|_2} - 5\cdot \textrm{SSIM}(\x_{\textrm{recon}}, \x_{\textrm{gt}})$, which leads to better image quality metrics in our experiments compared to using solely the normalized $\ell_2$ loss. The reconstruction loss function of the LOUPE model includes the unnormalized $\ell_2$ error to be consistent with the original work. 

For the alternating training framework, recall that we solve the optimization problem~\eqref{eqn::mb_loss} to provide labels (adaptive masks with respect to particular input) to train \mnet. We set the consistency parameter $\lambda=5\times10^{-4}$.  In practice, it is necessary to adjust the choice of the sparsity %implicit 
control parameter $\alpha$ with respect to different image inputs in the training process, 
%in order to maximize the exploitation of all training images, 
as an improper choice of $\alpha$ may let the \mb{} protocol to output degenerate masks (masks for multiple different objects being the same) or %make insufficient improvement over 
not improve the initial masks. 
Hence, we pre-specify an $\alpha$ grid and use a common $\alpha_0 = 2\times 10^{-5}$ as the initial setting with respect to each batch in the training set. If the returned masks from the \mb{} protocol fail to demonstrate sufficient adaptivity with respect to the given image batch or do not get updated from the input, we check the sampling ratio of the failed refined masks. If the failed refined masks have higher (lower) sampling ratio than the targeted value before they go through the sampling-budget control, we increase (decrease) $\alpha$ to the next value in the pre-assigned $\alpha$ grid and let the image batch to go through the \mb{} again. For $4\times$ acceleration of $k$-space, the $\alpha$-grid is chosen as a geometric sequence from $10^{-5.7}$ to $10^{-3.9}$ with quotient $10^{0.2}$, and for $8\times$ acceleration of $k$-space, it is a geometric sequence from $10^{-5.01}$ to $10^{-3.61}$ with quotient $10^{0.2}$. 

We use the RMSprop optimizer in PyTorch through all the training instances. In the alternating training framework, for each batch, we carry out $T=20 $ steps of iteration inside the \mb{} process to generate labels of adaptive masks with the learning rate $5\times 10^{-4}$ for Unet and $5\times 10^{-3}$ for mask characterization parameters $\boldsymbol{\xi}$. The learning rate for \mnet{} is $5\times10^{-4}$, and 40 steps of update are made to \mnet{} weights with respect to one batch of adaptive mask labels output by \mb{}. We run the alternating training framework for 10 epochs and obtain the corresponding \mnet{} and Unet-co-trained.

When training separate reconstructors (both Unet and MoDL), we used an initial learning rate $10^{-5}$ and the \texttt{ReduceLROnPlateau} scheduler in PyTorch to regulate the learning rate with patience 5, reduce factor 0.8 and minimal learning rate $10^{-6}$. Separate training of all reconstructors uses 40 epochs for each reconstructor.

When training the LOUPE framework, we use the learning rate $10^{-4}$ to update the mask parameters and $10^{-5}$ to update the Unet in the LOUPE framework. The slope parameter is set as 1. The LOUPE model is trained for 40 epochs. These settings lead to good performance of the LOUPE masks.

Regarding the sampling setting, in the 8$\times$ acceleration case, we set the base low-frequency initially observed lines to 8 rows, and remaining sampling budget for high-frequency information as 32 rows; and in the 4$\times$ acceleration case, the base low-frequency observation budget is 16 rows and the remaining sampling budget for high-frequency information is 64 rows.

\begin{figure}[!ht]
    \centering
    \begin{subfigure}[b]{.6\linewidth}
        \centering\captionsetup{width=\linewidth}
        \includegraphics[width=\linewidth]{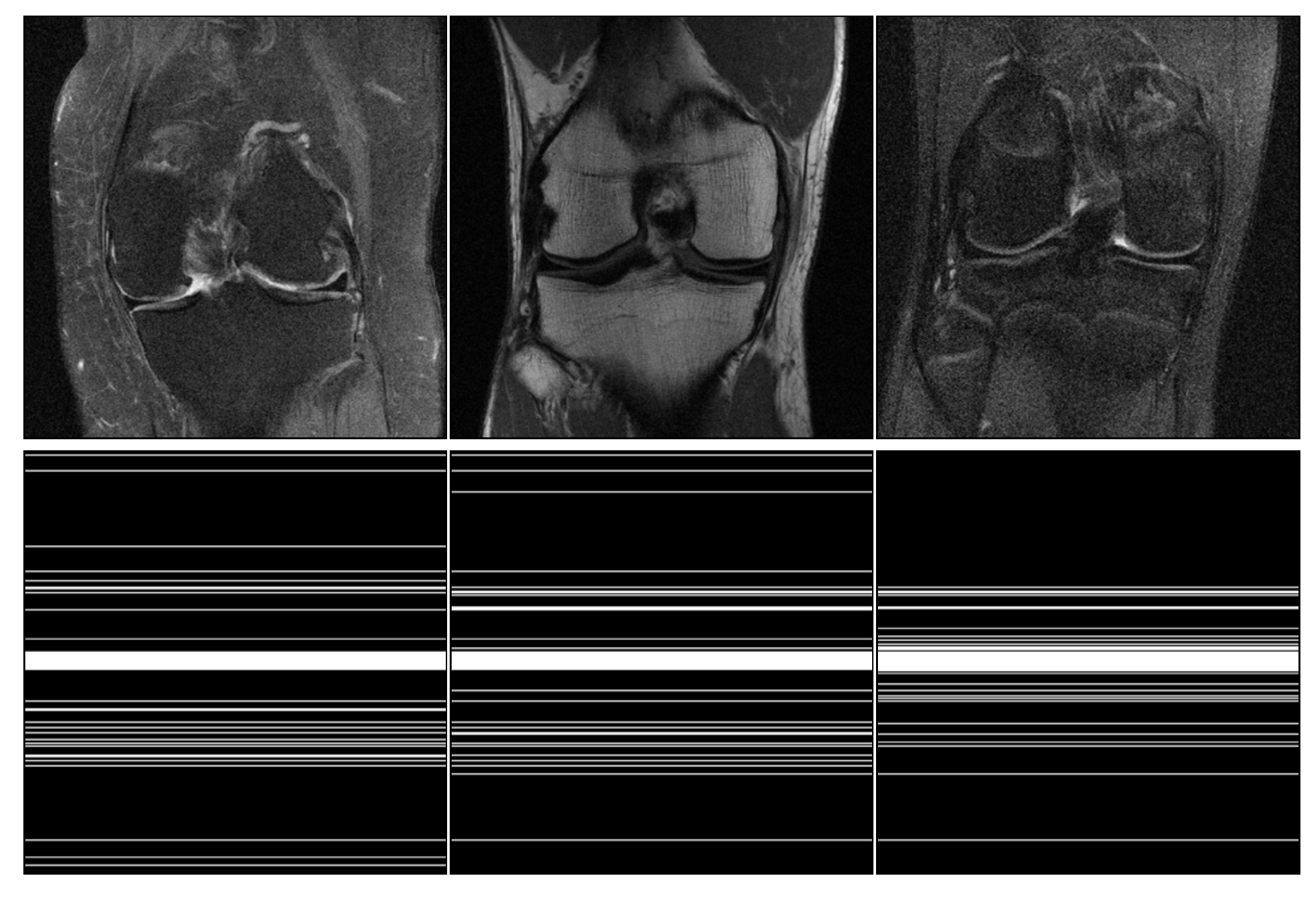}
        \label{fig:8x_mnet_mask_examples}
        \caption{8$\times$ acceleration masks (second row) predicted by \mnet{} for different training slices (first row): Base low-frequency information contains the central 8 rows, and the sampling budget for high-frequency information is 32 rows.}
    \end{subfigure}%
    \vfill
    \begin{subfigure}[b]{.6\linewidth}
        \centering\captionsetup{width=\linewidth}
        \includegraphics[width=\linewidth]{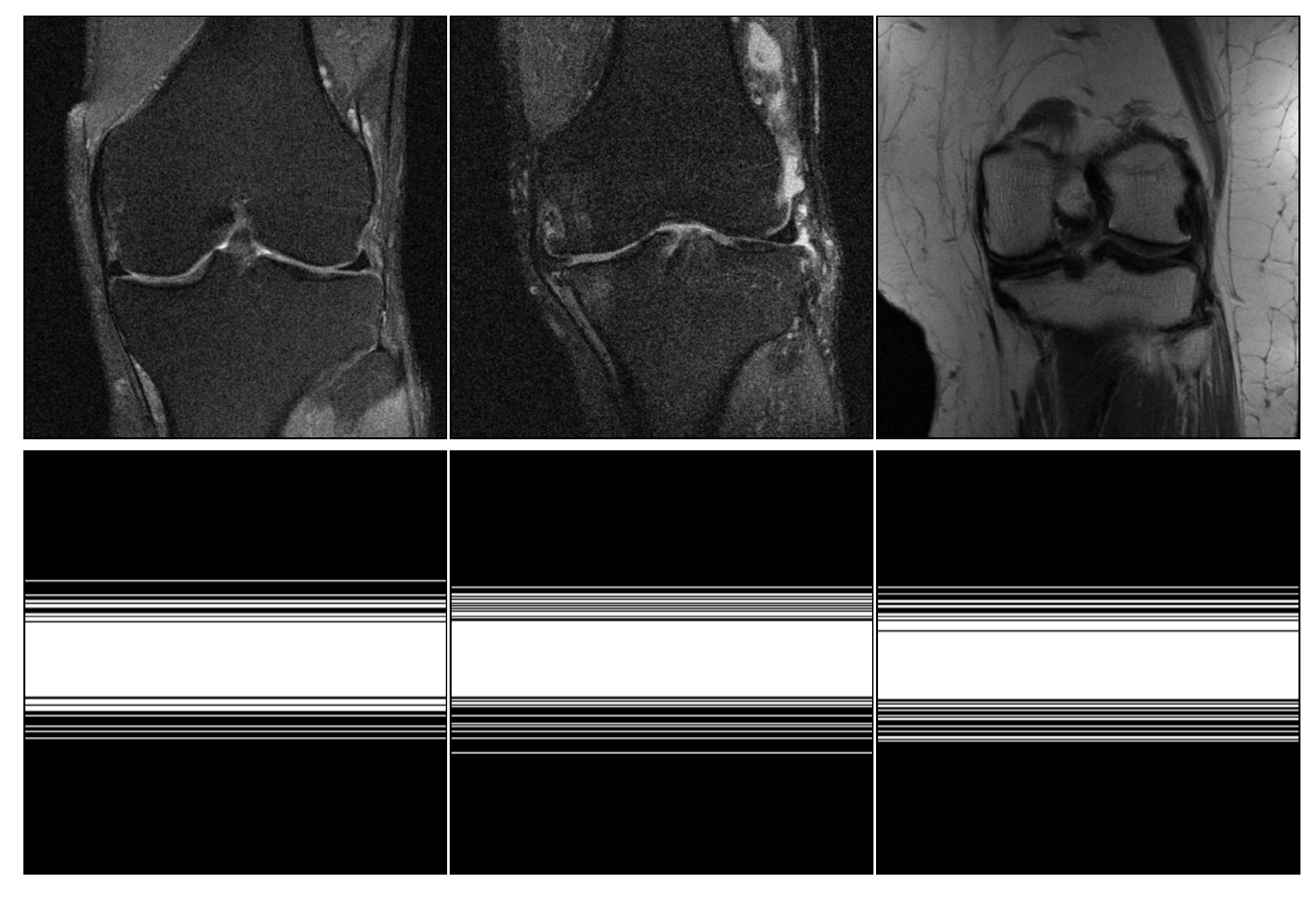}
        \label{fig:4x_mnet_mask_examples}
        \caption{4$\times$ acceleration masks (second row) predicted by \mnet{} for different training slices (first row). Base low-frequency information contains the central 16 rows, and the sampling budget for high-frequency information is 64 rows.}
    \end{subfigure}
   \caption{Adaptively predicted masks from \mnet{} for slice examples in the single-coil FastMRI knee dataset.}
  \label{fig::mnet_mask_examples}
\end{figure}

\subsection{Results}
In Figure~\ref{fig::mnet_mask_examples}, we visualize the adaptive masks output from the adaptive sampler \mnet{} trained by Algorithm~\ref{alg::alternating_training}. Knee images in Figure~\ref{fig::mnet_mask_examples} are the ground truth images. The \mnet{} sampler sees the low-frequency information collected in $k$-space and outputs the corresponding full subsampling patterns for sampling high-frequency information in a single pass, which are shown for each image and are object-adaptive.
%demonstrated below each knee image.

\begin{figure*}[!ht]
  \centering
    \includegraphics[width=.8\textwidth]{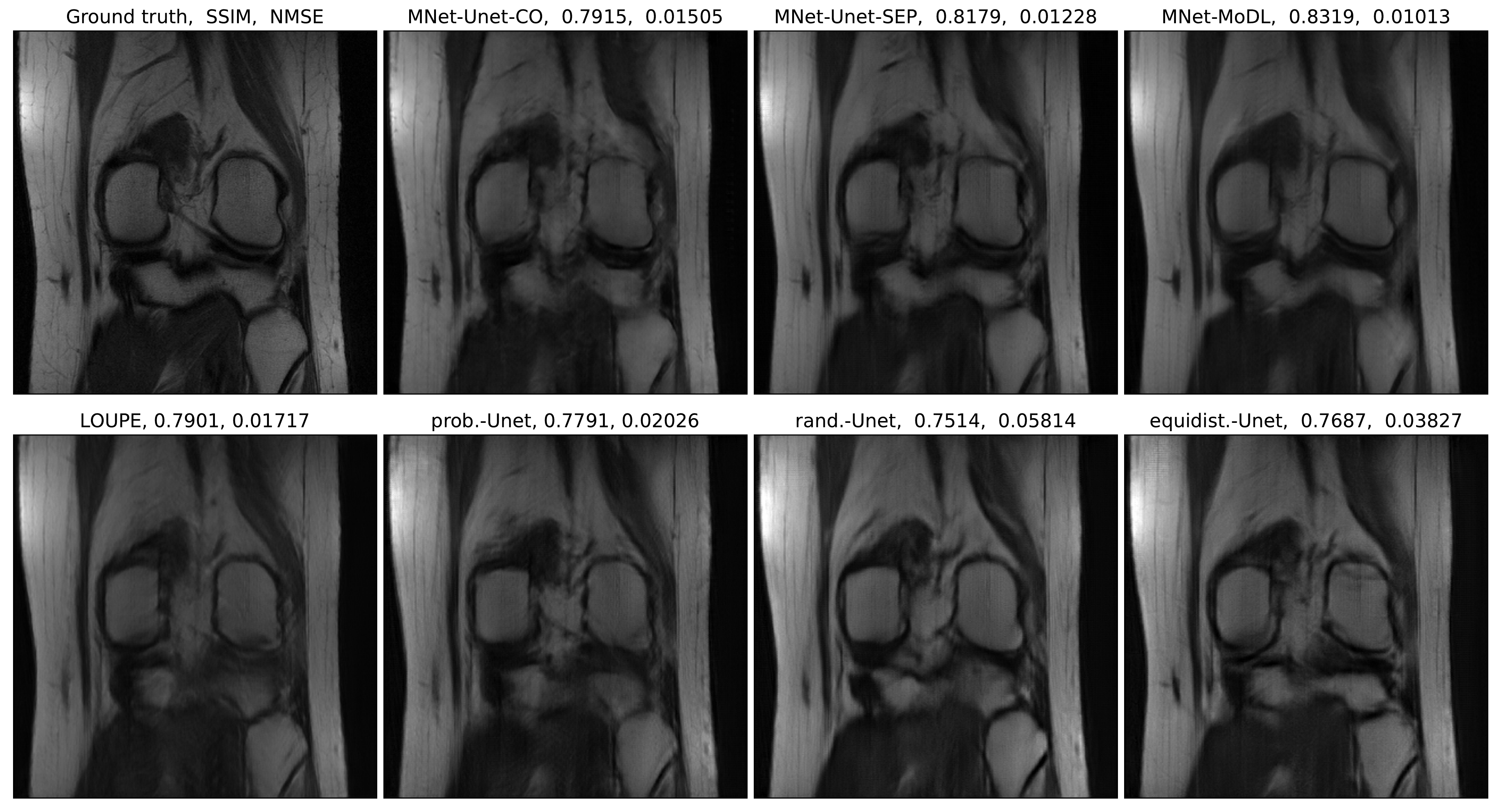}
  \caption{Reconstruction results from various combinations of sampler-reconstructor pairs with 8$\times$ acceleration of $k$-space.}
  \label{fig::8fold_recon_examples}
\end{figure*}

\begin{figure*}[!ht]
  \centering
  \includegraphics[width=.75\textwidth]{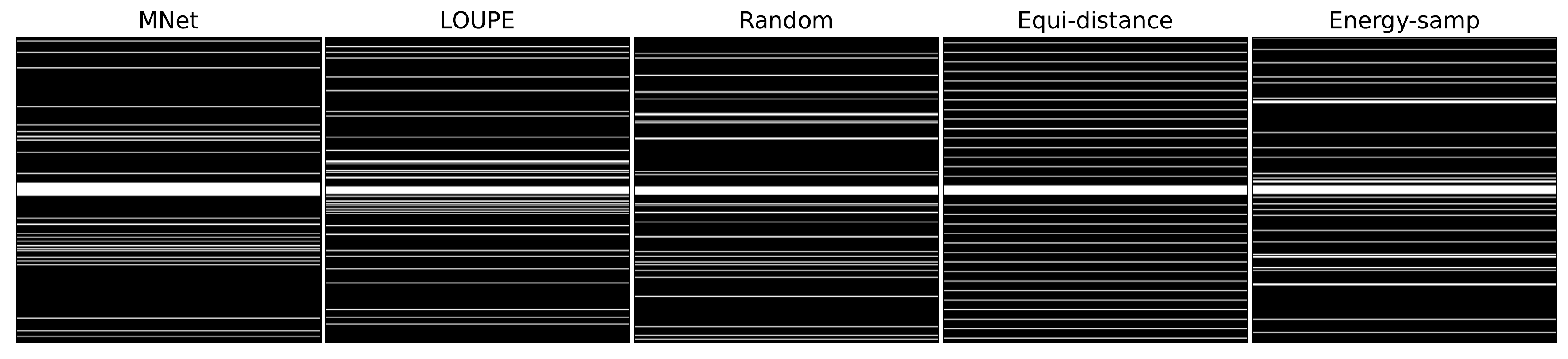}
  \caption{Comparison of different masks with respect to the object in Figure~\ref{fig::8fold_recon_examples} with $8\times$ acceleration.}
  \label{fig::8fold_mask_cmp}
\end{figure*}

Figure~\ref{fig::8fold_recon_examples} is an example showing the reconstructed images by several different reconstructors based on information collected from $k$-space according to different masks. The different corresponding masks are shown in Figure~\ref{fig::8fold_mask_cmp}. We consider the following combinations of sampler-reconstructor pairs: \mnet{} and co-trained Unet (Unet-CO), \mnet{} and follow-up separately-trained Unet (Unet-SEP), \mnet{} and a follow-up separately-trained MoDL reconstructor (\mnet{}-MoDL), LOUPE co-trained sampler and reconstructor, random sampling mask and separately-trained Unet (rand.-Unet), equi-distance mask and separately-trained Unet (equidist.-Unet), and energy density-based probabilistic mask and separately-trained Unet (prob.-Unet). We note the SSIM and NMSE values above the shown images.

One should note that the adaptive \mnet{} mask shown in Figure~\ref{fig::8fold_mask_cmp} is for the ground truth object in Figure~\ref{fig::8fold_recon_examples}, while the underlying probabilistic parametrization of LOUPE mask in Figure~\ref{fig::8fold_mask_cmp} is fixed after the training process completes with respect to the training dataset. The random mask is re-generated for each different input during the training and testing process, so the random mask used in the sampling process can differ from one object to another. The equi-distance mask is fixed with respect to the entire training and testing dataset given the pre-assigned amount of low frequencies observed and the amount of sampling budget for remaining high frequencies.

\begin{figure*}[!ht]
  \centering
  \includegraphics[width=.9\textwidth]{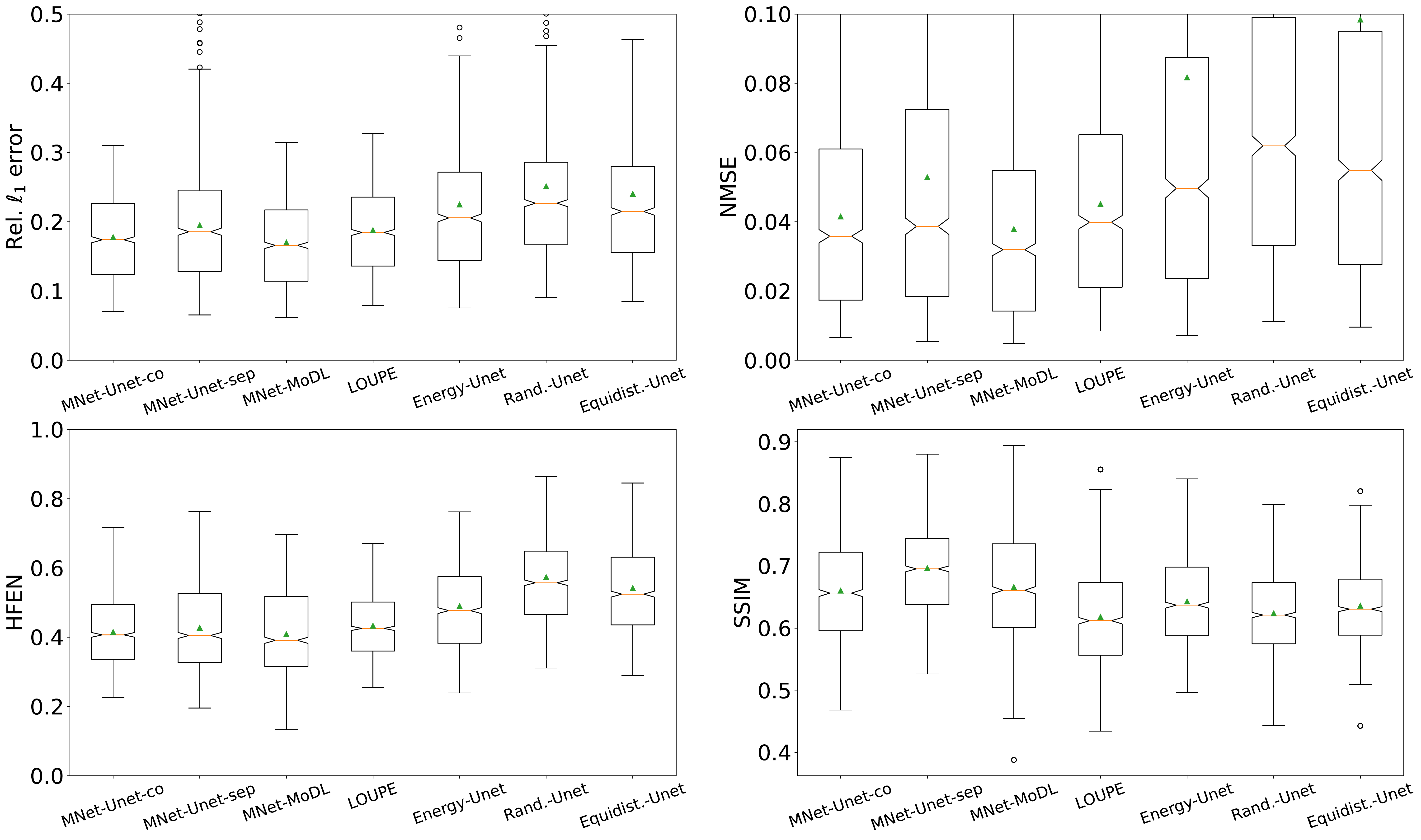}
  \caption{Comparison via box plots of reconstruction accuracy of different sampling-reconstructor combinations with $8\times$ acceleration of $k$-space. The triangle-shaped dot inside each box is the mean of the data in the box. Relative $\ell_1$ error is defined as $\frac{\|\x_{\textrm{recon}} - \x_{\textrm{gt}}\|_1}{\|\x_{\textrm{gt}}\|_1}$ for $\x_{\textrm{recon}}$ denoting the reconstruction and $\x_{\textrm{gt}}$ the ground truth image.}
  \vspace{0.1in}
  \label{fig::8fold_accuracy}
\end{figure*}

The complete reconstruction accuracy comparison for 8$\times$ acceleration is shown in Figure~\ref{fig::8fold_accuracy} using box plots. We consider four accuracy criteria to characterize reconstruction quality compared to the ground truth image: relative $\ell_1$ error (normalized mean absolute error, NMAE), normalized mean square error (NMSE), structural similarity index (SSIM), and high frequency error norm (HFEN) \cite{sai2011dlmri}.

%and relative $\ell_2$ error is $\frac{\|\x_{\textrm{recon}} - \x_{\textrm{gt}}\|_2}{\|\x_{\textrm{gt}}\|_2}$.

\begin{table}[!ht]
\centering
\begin{tabular}{ lccccc } 
\toprule
Method & NMAE & NMSE & HFEN & SSIM \\%  & PSNR\\
\midrule
\mnet{}-Unet-CO & 0.1778 & 0.0415 & 0.4150 & 0.6603 \\% & 27.3470  \\
\mnet{}-Unet-SEP & 0.1949 & 0.0529 & 0.4274 & \textbf{0.6965}\\% & 26.9146 \\
\mnet{}-MoDL & \textbf{0.1701} & \textbf{0.0379} & \textbf{0.4087} & 0.6662 \\% & \textbf{27.9473}\\
LOUPE & 0.1879 & 0.0451 & 0.4330 & 0.6180\\% & 26.8275\\
prob.-Unet & 0.2249 & 0.0817 & 0.4899 & 0.6429\\
rand.-Unet & 0.2512 & 0.1124 & 0.5735 & 0.6237\\% & 24.7051 \\
equidist.-Unet & 0.2403 & 0.0984 & 0.5418 & 0.6359\\% & 24.2393\\
\bottomrule
\end{tabular}
\caption{Reconstruction accuracy comparisons at 8$\times$ acceleration ratio between various sampler-reconstructor combinations. The values in the table are mean values of each box in Figure~\ref{fig::8fold_accuracy}.}
\label{table::8x_result}
\end{table}
Figure \ref{fig::8fold_accuracy} shows that advantages of different combination of samplers and reconstructors under different criteria. The \mnet-MoDL pair outperformed other combinations in terms of global accuracy (relative $\ell_1$ error and NMSE) and reconstruction quality of high frequency portions (HFEN), while the \mnet-Unet-SEP pair showed better performance in terms of feature characterization (SSIM). Although separately trained sophisticated unrolled-network reconstructor (MoDL) demonstrates higher reconstruction accuracy in multiple criteria, we highlight the comparable performance from the \mnet-Unet-CO pair due to the convenience of directly using a co-trained reconstructor without initiating a separate training effort. Our object-adaptive sampling approach through \mnet{} outperforms the LOUPE method, and both learning-based approaches are consistently better than the baseline methods that conduct random sampling and equi-distance sampling in $k$-space, respectively. 

The performances of the proposed sampler-reconstructor pairs characterized by NMSE and SSIM in this work are on par with the results reported in the FastMRI leaderboard.
Few outliers in the boxplots indicates better stability of co-trained Unet reconstructor and separately trained MoDL reconstructor compared to the separately trained Unet-reconstructor. 

One aspect of comparing these samplers is reproducibility. We point out that after the training is concluded, the \mnet{} mask prediction is deterministic and has reproducibility guarantee, while LOUPE methods and other probability-based samplers give varying output due to their stochastic nature. 
Our results indicate the importance of an object-adaptive sampler compared to population-adaptive samplers such as LOUPE and the energy density-based masks. Moreover, we observed that combining a random sampler with a sophisticated reconstructor MoDL does not bring as good reconstruction results as our method; i.e., making both the sampler and reconstructor adaptive offers benefits. 

\begin{figure*}[!ht]
  \centering
    \includegraphics[width=.9\textwidth]{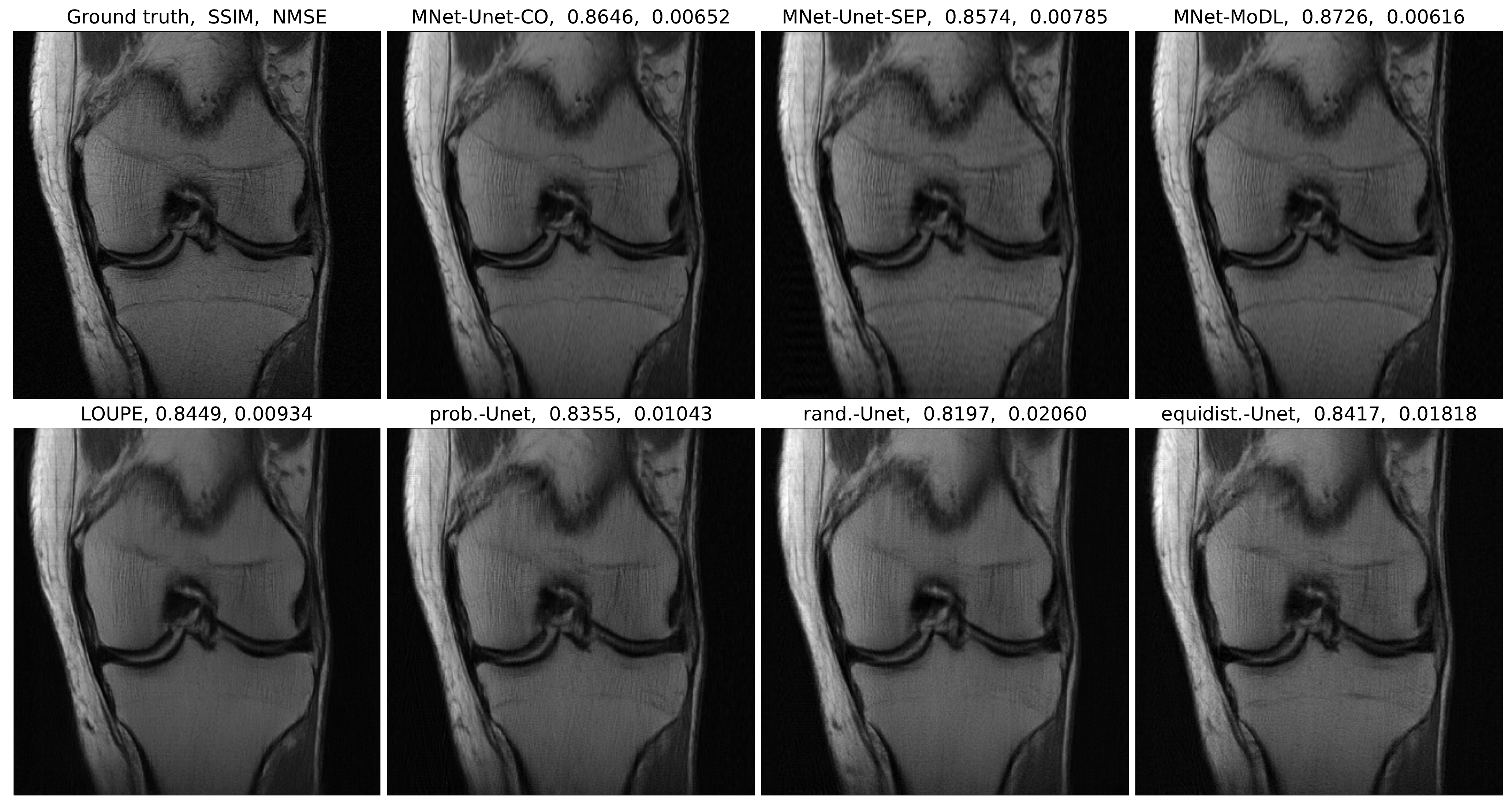}
    \caption{Reconstruction results from various combinations of sampler-reconstructor pairs with 4$\times$ acceleration of $k$-space.}
  \label{fig::4fold_recon_examples}
  \vspace{0.1in}
\end{figure*}

\begin{figure*}[!ht]
  \centering
  \includegraphics[width=.75\textwidth]{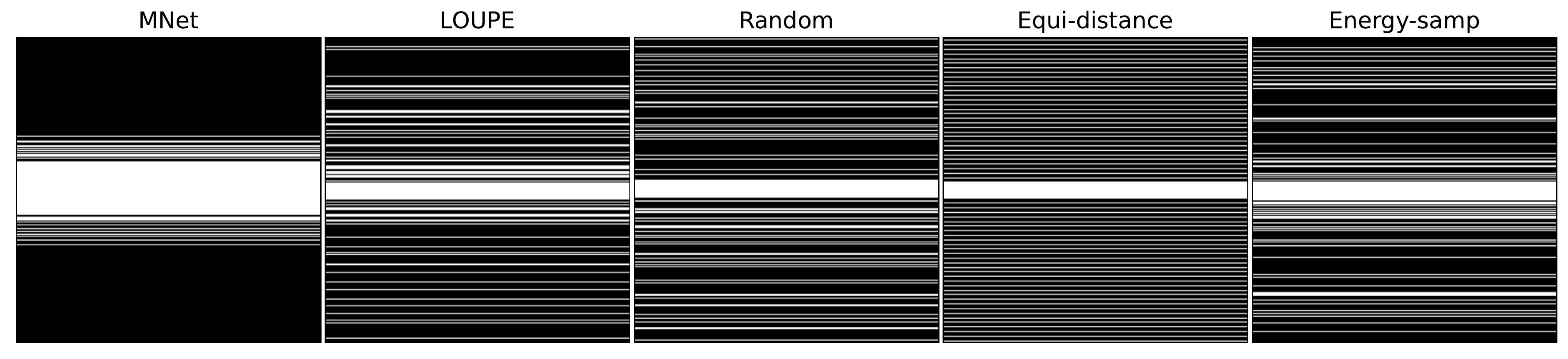}
  \caption{Comparison of different masks with respect to the object in Figure~\ref{fig::4fold_recon_examples} at $4\times$ acceleration.}
  \label{fig::4fold_mask_cmp}
\end{figure*}

\begin{table}[!ht]
\centering
\begin{tabular}{ lccccc } 
\toprule
Method & NMAE & NMSE & HFEN & SSIM \\% & PSNR\\
\midrule
\mnet{}-Unet-CO & 0.1480 & 0.0294 & 0.2673 & 0.7510 \\% & 29.3977  \\
\mnet{}-Unet-SEP  & 0.1654 & 0.0378 & 0.3113 & \textbf{0.7660} \\% & 28.5163 \\
\mnet{}-MoDL & \textbf{0.1437} & \textbf{0.0274} & \textbf{0.2663} & 0.7584 \\% & \textbf{29.7685}\\
LOUPE & 0.1540 & 0.0304 & 0.3356 & 0.7381 \\% & 28.8016\\
prob.-Unet & 0.1776 & 0.0449 & 0.3747 & 0.7520 \\
rand.-Unet & 0.2026 & 0.0646 & 0.4448 & 0.7348 \\% & 26.7310 \\
equidist.-Unet & 0.1897 & 0.0545 & 0.4132 & 0.7658 \\% & 26.3538\\
\bottomrule
\end{tabular}
\caption{Reconstruction accuracy comparisons at 4$\times$ acceleration for various sampler-reconstructor combinations. The values in the table are mean values of each box in Figure~\ref{fig::4fold_accuracy}.}
\label{table::4x_result}
\end{table}

\begin{figure*}[!ht]
  \centering
  \includegraphics[width=.8\textwidth]{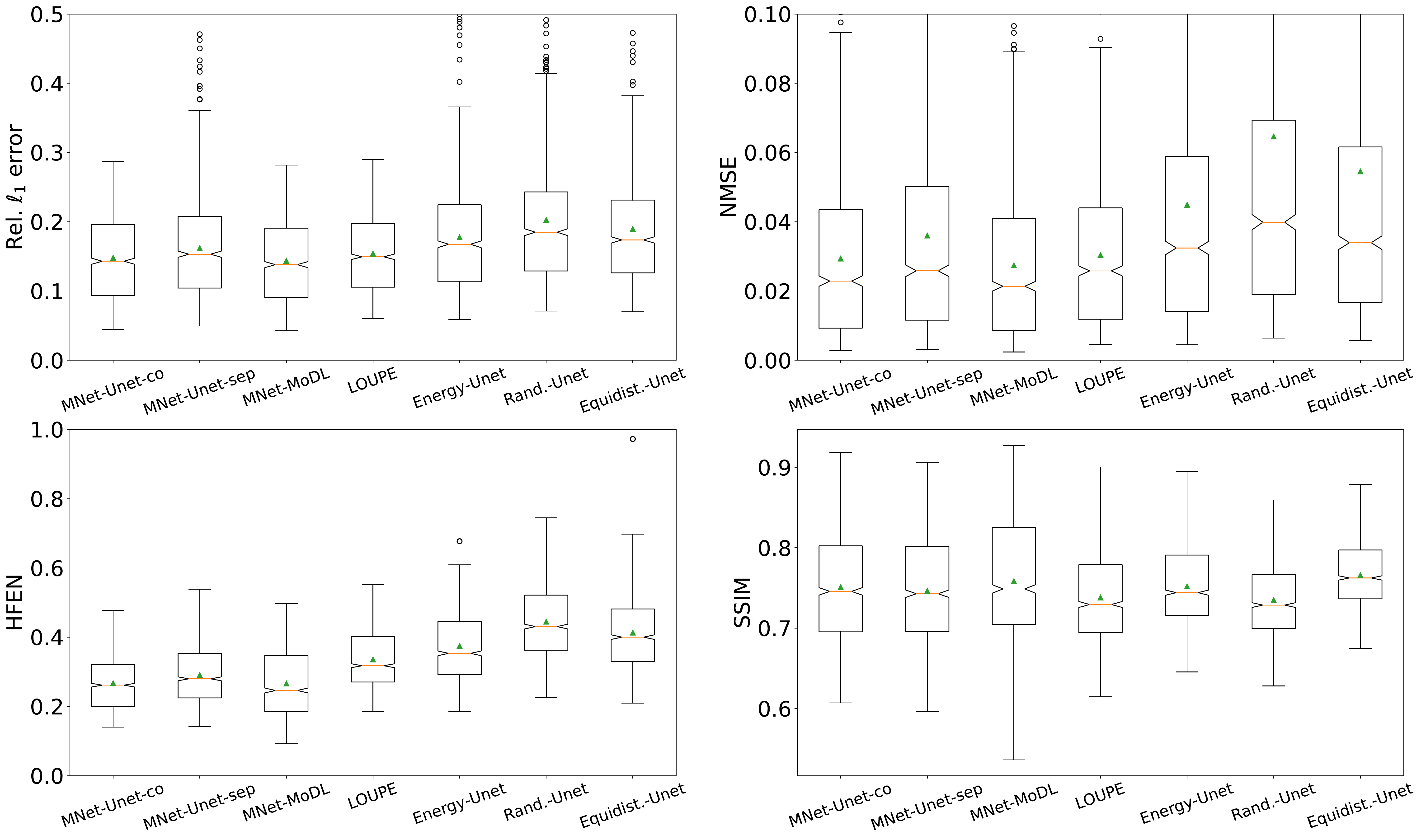}
  \caption{Comparison of reconstruction accuracy of different sampling-reconstructor combinations with $4\times$ acceleration of $k$-space.}
  \label{fig::4fold_accuracy}
\end{figure*}

Figures~\ref{fig::4fold_recon_examples}, \ref{fig::4fold_mask_cmp}, and  \ref{fig::4fold_accuracy} show the reconstruction examples, adaptive or compared mask examples, and reconstruction accuracy comparison results, respectively for the 4$\times$ acceleration case. The performance trend between methods in the 4$\times$ acceleration setting is similar to that for the 8$\times$ acceleration case discussed above. Tables \ref{table::8x_result} and \ref{table::4x_result} show the mean reconstruction accuracy comparison between various sampler-reconstructor combinations. 

We observe more concentration of the adaptive mask in the region of lower frequencies in the 4$\times$ acceleration setting. 
% This is due to the experiment setting that we start with more low frequencies in the 4$\times$ setting than in the 8$\times$ setting, and this creates a certain effect of frequency attraction. 
% It can also be due to specific hyperparameter settings for training in the 4$\times$ setting. 
We have also experimented on another 4$\times$ acceleration setting where we start with 8 lowest frequencies and sample 72 high frequencies. The mask output by \mnet{} has similar patterns as in Figure \ref{fig::4fold_mask_cmp}.

\section{Conclusions}
\label{sec::conclusions}
In this work, we propose an object-adaptive sampler for
undersampling $k$-space in MRI, while maintaining high quality of image reconstructions.
%sampling frequency information in the application of constructing MRI images. 
The sampler is realized by a convolutional neural network that takes as its input very limited $k$-space measurements (e.g., low-frequencies) and outputs the corresponding remaining sampling patterns at the desired sampling budget in a single pass.
%in the high-frequency domain. 
The training labels for the sampler network are generated internally in our framework using the \mb{} training protocol. We implemented the proposed sampler and alternating training framework on the single-coil knee FastMRI dataset and presented examples of adaptive masks with respect to various input image objects and undersampling factors. 
Our results show significant improvements in image quality with our single-pass object-adaptive sampling and reconstruction framework compared to other schemes.
%and shown numerical results characterizing the reconstruction quality.

\textbf{Future directions:} One central issue in all the learning-based adaptive sampling work is how to coordinate the parametrization of the mask and the challenge posed by the requisite mask binarization on gradient computation. 
We have resorted to numerical techniques in this work to address this issue. 
An alternative approach is to exploit delicate encoding schemes.  
We used the 1D line sampling in this work to demonstrate the capability of our framework, 
but it can be readily extended to subsampling phase encodes or shots in non-Cartesian settings by replacing the initial reconstructor with non-uniform FFT-based ones~\cite{fessler:07:onb}.
%while there is a large space to explore alternative sampling patterns such as non-Cartesian data acquisition or some particular type of viable sampling trajectories due to physical constraints as the prediction goal. 
To apply the \mnet{} method to predict general high-dimensional sampling patterns, some encoding schemes need to be introduced to reduce the complexity of %mask 
parametrization. For instance, the parametrization format in BJORK~\cite{BJORK_21} can be the target of \mnet{} output, or in other words, the \mnet{} predicts the interpolation coefficients $\mathbf{c}$ in BJORK, which can immediately render the 2D sampling trajectory to be object-specific. Meanwhile, developing training algorithms with some theoretical guarantees remains of importance for future work.

%While the learned \mnet{} can perform object-adaptive undersampling of Cartesian phase encodes, it can be readily extended to non-cartesian settings by replacing the initial reconstructor with non-uniform FFT-based ones~\cite{fessler:07:onb}.

In this work, an \mnet{} is trained with respect to fixed amount of initial (low-frequency) observations and fixed target sampling ratio. %in the high-frequency regime. 
The generalizability of such a trained \mnet{} to a different subsampling target remains to be investigated. Although modifying the sampling budget control procedure can automatically adjust the output of the trained \mnet{} to cater for different sampling ratio targets, it is unclear if the co-trained reconstructor can perform well with input images processed at a different subsampling ratio than that the reconstructor was trained for. 
This could be perhaps partially alleviated by re-training the reconstructor by itself with sampling patterns (post-binarization) output by \mnet{} with different sampling ratios than in training. Also, future work may explore how well a trained \mnet{} generalizes to data from different underlying distributions, and how one should quantify the correlation between various input images and characterize the corresponding influence on the predicted masks.

We also consider the setting of dynamic data acquisition to be of high interest for our future work. 
When the object evolves dynamically over time, it would be ideal to have a corresponding sampling scheme that copes with the changing object rather than directly applying the method developed for static MRI. For example, the input to the \mnet{} could include data from the previous frame(s).
%When the status of the testing object undergoes certain evolution according to some dynamics, although not of high volatility, it will be ideal to have a corresponding sampling scheme to cope with the changing object rather than directly applying the method developed for static setting.
Similarly, one can replace various terms in the \mb{} objective function~\eqref{eqn::mb_loss} with domain-knowledge motivated priors or other penalty terms (e.g., reconstruction errors or contrast in regions of interest)
that may boost the performance of the reconstructors to better capture certain features in the underlying data. 
%training dataset accordingly. 
Finally, while our experiments focused on single-coil data, the proposed \mnet{} method can directly generalize to the multi-coil setting, as one simply needs to revise the input format for the networks without changing the rest of the training framework. 

\section*{Acknowledgments}
We thank Dr. Michael McCann at Los Alamos National Laboratory, and Shijun Liang and Siddhant Gautam at Michigan State University for helpful discussions. We thank all reviewers for their constructive comments on our initial draft.

\bibliographystyle{alpha}
\bibliography{references}

\end{document}